\documentclass[aps,prb,twocolumn,groupedaddress,showpacs]{revtex4}
\usepackage{epsfig}
\begin{document}

\title{Ring Exchange and Phase Separation in the two-dimensional Boson 
Hubbard model}
\author{V.G. Rousseau, R.T. Scalettar}
\affiliation{Physics Department, University of California, Davis, California 95616, USA}
\author{G.G. Batrouni}
\affiliation{Institut Non Lin\'eaire de Nice, 1361 route des Lucioles, 06560 Valbonne, France}

\begin{abstract}
We present Quantum Monte Carlo simulations of the soft-core bosonic
Hubbard model with a ring exchange term $K$.  For values of $K$ which
exceed roughly half the on-site repulsion $U$, the
density is a non-monotonic function of the chemical potential,
indicating that the system has a tendency to phase separate.  This
behavior is confirmed by an examination of the density-density
structure factor at small momenta and real space images of the boson
configurations.  Adding a near-neighbor repulsion can
compete with phase separation, but still does not give rise to a stable
normal Bose liquid.
\end{abstract}

\maketitle

\section*{Introduction}

The most commonly studied Hamiltonians of quantum spins and bosons
include kinetic energy in the form of two site exchange.  However, it
was realized early on that, for lighter atoms, larger numbers of
particles can permute their positions, leading to ring exchange
terms.\cite{Thouless65} Interest in such models subsequently developed
for a number of reasons, most recently as a result of the possibility
of the existence of a normal bose liquid. Such a phase which might
reconcile apparently contradictory observations in cuprate
superconductors in the pseudogap regime where there is evidence of
local pairing without superconductivity.\cite{Paramekanti02,Doniach03}
Indeed, the magnitude of ring exchange terms in high-$T_c$ materials
could be a significant fraction of the two site
exchange.\cite{Coldea01,Katanin02,MullerHartmann02}

Over the past several years,
considerable numerical effort has been expended in looking at the
low temperature phases of
Hamiltonians including ring exchange.\cite{Melko04,Park02,Rousseau04,Roux05}
Interestingly,
while the normal bose liquid has remained somewhat elusive,
a rich panoply of other behavior has been observed, including
striped order, Neel antiferromagnetism, valence bond solids,
and phase separation.
The nature of the phase diagram and of the phase transitions
between the different ordered phases remains an area
of active inquiry.\cite{Sandvik05}
Very recently it was proposed\cite{Buchler05} that ring exchange terms might
be engineered into cold atomic gases, opening up the possibility
that the effects of such terms could be studied in a situation
where they can be tuned systematically, and hence the
theoretical predictions checked experimentally.

In this paper we will extend our previous work\cite{Rousseau04}
on phase separation
in the boson Hubbard model with ring exchange.
We focus on two dimensions, since that is the
lowest dimension allowing ring exchange processes, and it is the
effective dimension of the CuO$_2$ sheets of the cuprate
superconductors which have motivated much of the recent study of Bose liquid
phases.  Our key conclusion is that when the ring exchange energy scale
exceeds approximately one half the on-site repulsion,
the ground state is thermodynamically unstable to phase separation.
Adding a near-neighbor repulsion can prevent phase separation, but
the ground state always has either nonzero checkerboard or
superfluid order.  Because we use a break-up of the Hamiltonian
in constructing the path integral which has not been much
employed before, we also describe some of the technical
details of the simulation, including the effectiveness of different
local Quantum Monte Carlo (QMC) moves.

The most simple boson Hubbard model\cite{Fisher89} is:
\begin{equation}
  \label{Hamiltonian}
\hat\mathcal H=-t\sum_{\bigl\langle ij\bigr\rangle}
       (a_j^\dagger a_i + a_i^\dagger a_j)
+U \sum_{i} \hat n_i (\hat n_i -1)
\end{equation}
The operators $a_j^\dagger,a_j$ create (destroy) a boson on site $j$,
and obey commutation rules $\bigl[a_i,a_j^\dagger\bigr]=\delta_{ij}$.
The number operator is $\hat n_j = a_j^\dagger a_j$.  The hopping parameter
$t$ measures the kinetic energy and $U$ the strength of the on-site
repulsion.  The sum $\langle ij \rangle$ is over near neighbors on a
square lattice.  In the limit $U \rightarrow \infty$ this model maps
onto the spin-1/2 XY model, with the magnetization playing the role of
the bosonic density.  By now, the $T=0$ phase diagram of the boson
Hubbard Hamiltonian is well
known.\cite{Fisher89,Batrouni90,Kuhner00,Freericks96,Krauth91,vanOtterlo94,Prokofev04}
Away from commensurate
filling, the ground state is superfluid.  At commensurate fillings,
and for sufficiently large $U$, a Mott insulator forms in which each
site is occupied by an integer number of particles.\\

Our goal is to consider the effect of a ring exchange term:
\begin{equation}
\label{RingExchangeTerm}
\hat\mathcal K=-K\sum_{\left\lbrace p\right\rbrace}
       (a_1a_2^\dagger a_3^\dagger a_4+a_1^\dagger a_2a_3a_4^\dagger)
\end{equation}
The ring exchange term acts on four
site plaquettes
$\bigl\lbrace p\bigr\rbrace$,
destroying two particles which lie along one diagonal,
and creating them on the other.  Because of
the minus sign in (\ref{RingExchangeTerm}), the energy decreases when
the particles are exchanged.
The basic qualitative picture behind the suggestion that the ring
exchange term might give rise to a normal Bose liquid is that if one
starts with a superfluid phase, $K$ introduces local vortices in which
two particles jump in opposite directions.  These local twists, if
sufficiently favored energetically, could compete with, and ultimately
prevent, a coherent long range superflow of particles.  At the same
time, such a hopping process increases (local) quantum fluctuations,
and would not be expected to make the system incompressible (a Mott
insulator).  Therefore one might have a normal Bose liquid.

As we shall see, we find that this ring exchange term causes phase separation
rather than a normal Bose liquid.  It is natural to attempt to resurrect the liquid
by adding a longer range repulsion which acts against particle clustering.
The most simple form is a near-neighbor repulsion:
\begin{equation}
  \label{IntersiteTerm}
\hat\mathcal V=V\sum_{\bigl\langle ij\bigr\rangle} \hat n_i\hat n_j
\end{equation}
The effects of near neighbor repulsion in the absence of ring exchange
have been well-studied in the context of the boson Hubbard
Hamiltonian.\cite{Niyaz93} They are perhaps most simply ascertained by
considering the hard-core limit and the equivalent spin model.  As
commented earlier, the boson kinetic energy corresponds to an $XY$
spin coupling.  $V$ corresponds to a $Z$ coupling.  Thus when $V < 2t$
the spin model is in the $XY$ universality class (a boson superfluid)
and for $V > 2t$ the spin model is in the Ising universality class (a
checkerboard charge density boson insulator).  A considerable amount
of work has also been done on looking for supersolid phases which
combine superfluid and charge density wave (CDW)
order.\cite{OtherSS,Batrouni00}

\begin{figure}[t]
  \centerline{\epsfig{figure=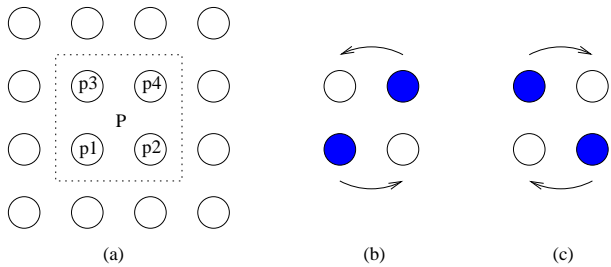,width=8cm}}
  \caption{\label{RingExchange}        {The ring exchange term
acts on each square plaquette of our 2d lattice (a), and allows the
exchange of two particles
from one diagonal (b) to the other (c).}}
\end{figure}

\section*{QMC:  The Partition Function as a Path Integral}

Our Quantum Monte Carlo simulations are performed using a World Line
algorithm with a decomposition involving four site matrix elements.
We discuss the algorithm in detail here
because this decomposition, which is similar but not identical to reference
\cite{Loh85}, is used considerably less frequently than the more
conventional approach which involves two site matrix
elements.\cite{twositedecoupling} In fact, to our knowledge, it has
not been employed before for Hamiltonians including ring exchange,
although stochastic series expansion approaches\cite{Sandvik91} are
formally very similar.  A specific issue we will address is the effect
of different local moves on the measurement of the energy.

We begin by representing the partition function as a path integral, in
which $e^{-\beta\hat\mathcal H}$ is the imaginary time evolution
operator, $0 \leq \tau \leq \beta$.  The goal is to take the
exponential of the full Hamiltonian, which cannot be computed, and
express it in terms of exponentials whose numerical values can be
written down.  In order to accomplish this, we write the full
imaginary time evolution operator as a product of infinitesimal
evolution operators over short imaginary times $\Delta \tau$:
\begin{equation}
  \label{Partition-1}
  \mathcal Z=\textrm{Tr }e^{-\beta\hat\mathcal H}=\textrm{Tr }
(e^{-\Delta \tau\hat\mathcal H})^M \hspace{1cm} \tau=\frac{\beta}M
\end{equation}
Here $M$ is chosen to be a sufficiently large integer so as to make
$(\Delta \tau)^2$ times any two of the energy scales in $\hat\mathcal H$ small, as
discussed further below.

We then divide the sum over all plaquette operators into four classes
depicted by the different shadings (colors on-line)
in Fig.~\ref{Decomposition}.  That
is, the full Hamiltonian is written as
\begin{equation}
  \hat\mathcal H=\hat\mathcal H_{1}+\hat\mathcal H_{2}+\hat\mathcal H_{3}+\hat\mathcal H_{4}
\end{equation}
where
\begin{equation}
  \hat\mathcal H_{n}=\sum_{p \in n}\hat\mathcal H_p
\end{equation}
groups together all plaquettes of a given shading.
In order to avoid overcounting the kinetic energy, on-site, and near
neighbor interaction terms in the Hamiltonian we define,
\begin{eqnarray}
   && \hat\mathcal H_p=\hat\mathcal K_p+\frac12     (\hat\mathcal T_p+
 \hat\mathcal V_p      )+\frac14\hat\mathcal U_p
\end{eqnarray}
with
\begin{eqnarray}
  && \hat\mathcal T_p=-t (a_{p1}^\dagger a_{p2}+a_{p3}^\dagger a_{p4}+
a_{p1}^\dagger a_{p3}+a_{p2}^\dagger a_{p4}+h.c.  ) \nonumber \\ &&
\hat\mathcal K_p=-K (a_{p1}a_{p2}^\dagger a_{p3}^\dagger a_{p4}+h.c.
) \nonumber \\ && \hat\mathcal U_p=U [\hat n_{p1} (\hat n_{p1}-1 )+
\hat n_{p2} (\hat n_{p2}-1 )  \nonumber \\ && \hspace{0.45cm}+\hat
n_{p3} (\hat n_{p3}-1 )+ \hat n_{p4} (\hat n_{p4}-1 ) ] \nonumber \\
&& \hat\mathcal V_p=V (\hat n_{p1}\hat n_{p2}+ \hat n_{p3}\hat
n_{p4}+\hat n_{p1}\hat n_{p3}+\hat n_{p2}\hat n_{p4} ).
\end{eqnarray}
It is important to emphasize that while plaquette operators acting on
neighboring plaquettes do not commute, all plaqette operators within
the same shading do commute, since they do not ``touch."  This
independence will enable us to compute their exponentials.

\begin{figure}[t]
   \centerline{\epsfig{figure=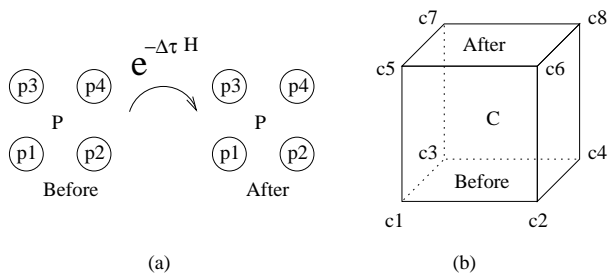,width=8cm}}
  \caption{\label{Plaquette} { The application of $e^{-\Delta \tau\hat\mathcal H}$
propagates the boson occupations on the four spatial sites of the
plaquette (a) in the imaginary time direction, and can be thought of
geometrically as converting each spatial plaquette into a cube (b)
whose bottom face contains the boson occupations before the action of
the operator, and whose top face contains the boson occupations after
the action of the operator.}}
\end{figure}

Now we express each infinitesimal evolution operator as a product:
\begin{eqnarray}
  \label{Partition-2}
  \nonumber && \hspace{-0.5cm} e^{-\Delta \tau\hat\mathcal
  H}=e^{-\Delta \tau (\hat\mathcal H_{1}+\hat\mathcal
  H_{2}+\hat\mathcal H_{3}+\hat\mathcal H_{4} )}\\ &&
  \hspace{-0.2cm}\approx e^{-\Delta \tau\hat\mathcal
  H_{1}}e^{-\Delta \tau\hat\mathcal H_{2}}e^{-\Delta
  \tau\hat\mathcal H_{3}}e^{-\Delta \tau\hat\mathcal H_{4}}
\end{eqnarray}
Substituting expression (\ref{Partition-2})
in expression (\ref{Partition-1}), we get:
\begin{equation}
  \label{Partition-3}
  \hspace{-0.2cm}\mathcal Z=\textrm{Tr} ( e^{-\Delta \tau\hat\mathcal
H_{4}}e^{-\Delta \tau\hat\mathcal H_{3}} e^{-\Delta
\tau\hat\mathcal H_{2}}e^{-\Delta \tau\hat\mathcal H_{1}})^M
\end{equation}

The errors made in breaking up $e^{-\Delta \tau \hat \mathcal H}$,
Eq.~\ref{Partition-2}, go as the commutator of the individual pieces,
and therefore as $(\Delta \tau)^2$.  It might be thought that the
accumulation of the $(\Delta \tau)^2$ errors across the $M$ imaginary
time slices might lead to an error linear in $\Delta \tau$.  However,
the expectation value of any Hermitian operator\cite{Fye86,Fye87}
has a Trotter
error which is quadratic in $\Delta \tau$.

\begin{figure}[h!]
   \centerline{\epsfig{figure=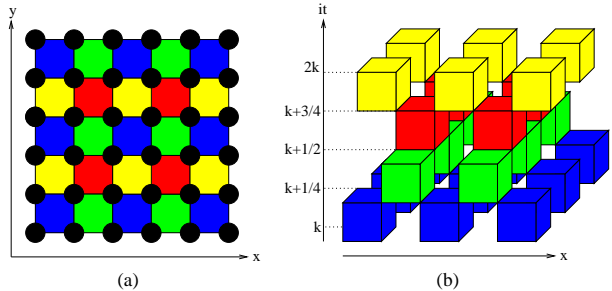,width=8cm}}
  \caption{\label{Decomposition} {(a) Plaquettes with same shading
(color online) are
independent. (b) Introducing a dimension in imaginary time makes the
plaquettes become cubes, each representing an infinitesimal evolution
operator. Cubes with same shading don't touch each other, while cubes
with different shading act at different times and therefore are
decoupled.}}
\end{figure}

To complete the evaluation of the trace, we work in the basis of
occupation numbers $\left|\left\lbrace n\right\rbrace\right\rangle$.
Inserting a complete set of states $I=\sum_{\left\lbrace
n\right\rbrace}\left|\left\lbrace
n\right\rbrace\right\rangle\left\langle\lbrace n\rbrace\right|$
between each pair of exponentials, we get:
\begin{equation}
  \label{Partition-4}
  \begin{array}{ll}
    \mathcal Z=\displaystyle\sum_{\left\lbrace
    n^k\right\rbrace}&\displaystyle{\prod_{k=1}^M}\left\langle\left\lbrace
    n^{k+1}\right\rbrace\right|e^{-\Delta \tau\hat\mathcal
    H_{4}}\bigl|\bigl\lbrace n^{k+3/4}\bigr\rbrace\bigr\rangle\\
    &\times\left\langle\left\lbrace
    n^{k+3/4}\right\rbrace\right|e^{-\Delta \tau\hat\mathcal
    H_{3}}\left|\left\lbrace n^{k+1/2}\right\rbrace\right\rangle\\
    &\times\left\langle\left\lbrace
    n^{k+1/2}\right\rbrace\right|e^{-\Delta \tau\hat\mathcal
    H_{2}}\left|\left\lbrace n^{k+1/4}\right\rbrace\right\rangle\\
    &\times\left\langle\left\lbrace n^{k+1/4}
    \right\rbrace\right|e^{-\Delta \tau\hat\mathcal
    H_{1}}\left|\left\lbrace n^k\right\rbrace\right\rangle
  \end{array}
\end{equation}

The superscripts $k$ appearing in the sets $\left\lbrace
n^k\right\rbrace$ label the imaginary time, emphasizing that a
different complete set of states has been introduced between each pair
of exponentials. Since operators acting on plaquettes with the same
shading commute, each matrix element in (\ref{Partition-4}) can be
written as a product of four site matrix elements
\begin{equation}
  \!\bigl\langle\bigl\lbrace n^{q+1/4}\bigr\rbrace\bigr|e^{-\Delta
  \tau\hat\mathcal H_{n}}\left|\left\lbrace
  n^q\right\rbrace\right\rangle\!=\!\!\prod_{p(q)}\!\bigl\langle\mathcal
  P^{q+1/4}\bigr|e^{-\Delta \tau\hat\mathcal H_p}\left|\mathcal
  P^q\right\rangle
\end{equation}
where $\mathcal P^q$ represents the state at imaginary time $q$ ($q$
is a multiple of $\frac 14$) of the four sites belonging to plaquette
$p$, and $\prod_{p(q)}$ is a product over all matrix elements between
times $q$ and $q+1/4$.
The entire evolution is represented by an interconnected set
of the cubes of Fig.~\ref{Plaquette}b,
as illustrated in Fig.~\ref{Decomposition}b.

Since each $\hat\mathcal H_p$ conserves the sum of the boson numbers on the four
spatial sites, the occupation numbers living on these cubes trace out
continous `world-lines' of particles which wiggle and deform during
their evolution.  Furthermore, because we are evaluating the trace of
$e^{-\beta \hat\mathcal H}$, the occupation numbers and associated world
lines are periodic in imaginary time.  We will discuss the sampling of
the expression for $\mathcal Z$, and the associated conservation laws, in more
detail below.

Summarizing, the partition function takes the simple form of a sum of
products of four site matrix elements,
\begin{equation}
  \label{Partition-5}
  \mathcal Z=\sum_{\left\lbrace\mathcal P^k\right\rbrace}\prod_{k=1}^M
\prod_{p(k)}\bigl\langle\mathcal P^{k+1/4}\bigr|e^{-\Delta \tau\hat\mathcal H_p}\left|\mathcal
P^k\right\rangle
\end{equation}
The imaginary time index $k$ runs from $1$ to $M$ with step $\frac
14$. Each matrix element in (\ref{Partition-5}) can be computed by
numerical diagonalization if we assume that occupation numbers are
never greater than four.  (This restriction is satisfied in all
simulations here, since the average densities studied are of order
unity.\cite{highocc}) Each site has five possible states and the
number of different states for one plaquette is $5^4=625$.  To compute
the exponential of $\hat\mathcal H_p$ then requires the diagonalization of a
$625\times 625$ matrix.

\section*{Sampling of the partition function}

\begin{figure}[h!]
   \centerline{\epsfig{figure=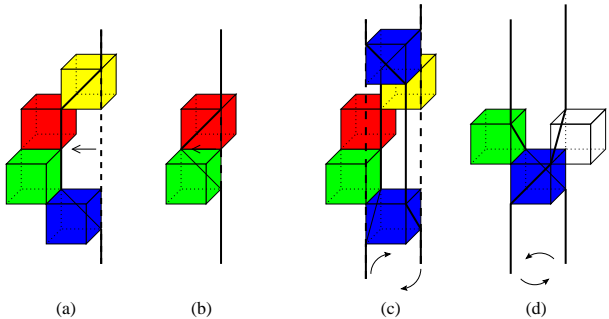,width=8cm}}
  \caption{\label{LocalMoves} {To ensure ergodicity of the algorithm,
four kinds of moves must be included: (a) pull moves, (b) ``baby''
pulls, (c) twist moves and (d) ``baby'' twist moves.}}
\end{figure}

In order to perform measurements, one needs to generate sets of world
lines with the Boltzmann weight given by the summand of
Eq.~\ref{Partition-5}. This is done using a Metropolis algorithm. One
starts with straight world lines, then suggests local deformations and
accepts them with probability $p=\textrm{min}(1,{\cal R})$, where
${\cal R}$ is the ratio of the Boltzmann weight after the move to that
before.  In the usual way, this satisfies detailed balance.  To ensure
ergodicity of the algorithm, different kinds of moves have to be
included, like pull moves (Fig.~\ref{LocalMoves}a), ``baby'' pull
moves (Fig.~\ref{LocalMoves}b), twist moves (Fig.~\ref{LocalMoves}c),
and ``baby'' twist moves (Fig.~\ref{LocalMoves}d).  Only a few of the
plaquette matrix elements change when one of these local moves is
performed.  The most complex is the twist move
(Fig.~\ref{LocalMoves}c) which involves 10 matrix elements (only 5
cubes are shown in order to make the figure clearer).

\begin{figure}[h!]
   \centerline{\epsfig{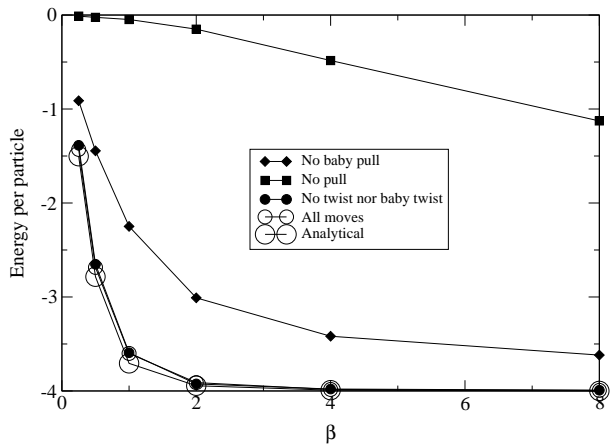}}
  \caption{\label{EnergyVsMoves-1} Energy per particle for
  free bosons as a function of the inverse temperature
  $\beta$ with and without including pull and baby pull moves.}
\end{figure}

These different kinds of moves are required in order to make the
algorithm ergodic.  As an illustration, we show in
Fig.~\ref{EnergyVsMoves-1} the energy per particle of free bosons as a
function of the inverse temperature $\beta$, with various types of
moves suppressed. We can see that the correct energy $E(\beta)$ is
obtained only when both pull and ``baby'' pull moves are included.

\vskip 0.5cm

\begin{figure}[h!]
   \centerline{\epsfig{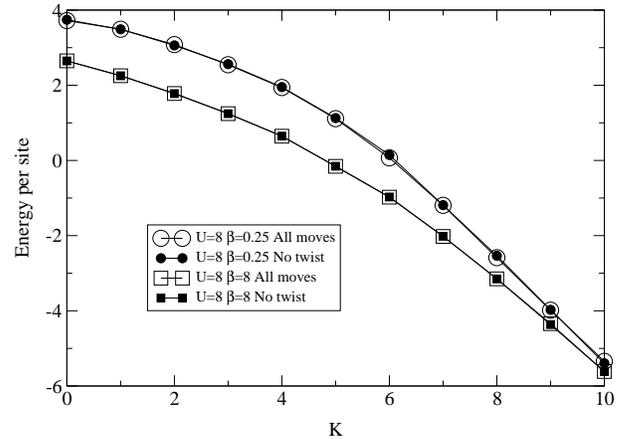}}
  \caption{\label{EnergyVsMoves-2} Energy per particle for
  interacting bosons as a function of the ring exchange strength
  $K$ with and without including twist moves.}
\end{figure}

It is notable in Fig.~5 that the correct energy is obtained even without the
twist moves.
It might be thought that this is because $K=U=0$ and that
twist moves will become important when
turning on interactions, especially with ring exchange
processes. However this is not the case, as can be seen in
Fig.~\ref{EnergyVsMoves-2}.  We believe this to be an advantage of the
four site decomposition over the two site one.  The four site
decomposition already analytically includes twists in the evaluation
of the matrix elements of the exponentials of the $H_n$.

While including only pull and ``baby'' pull moves seems to get the energy right on the lattice
sizes studied above, our simulations are done with all moves included.
These local moves all take the form of local distortions of existing
world-lines and hence cannot change $\rho$.
Thus our code works in
the canonical ensemble.  It simulates whatever value of $\rho$ with
which we initialize the lattice.  If desired, we can make contact with
the grand-canonical ensemble, by {\it computing} the chemical
potential as the energy cost, in the ground state, of adding a boson
to the system, $\mu(N_b)=E(N_b+1)-E(N_b)$.  Actually, canonical
simulations have certain advantages in exploring phase separation, as
we shall see, and as has previously been described.\cite{Batrouni00}

\begin{figure}[h!]
   \centerline{\epsfig{figure=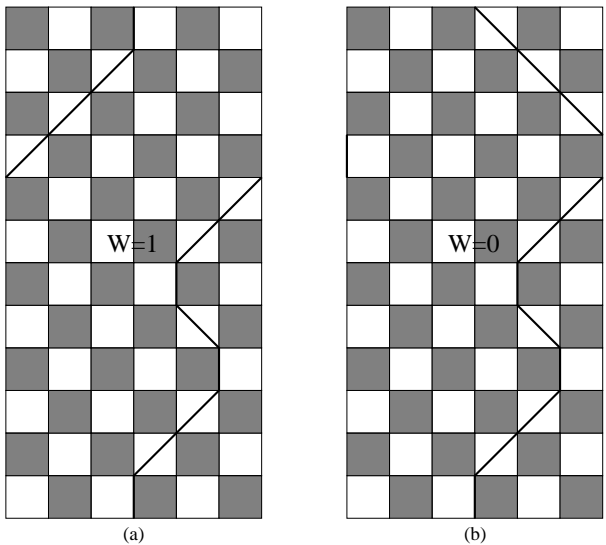,width=8cm}}
  \caption{\label{Winding}        {Example of world line with $W=1$ (a)
and $W=0$ (b) winding numbers in
  the case of a one dimensional lattice. Configurations with non-zero
winding number cannot be reached with local moves.}}
\end{figure}

Our simulations also are confined to the zero-winding sector of the
space of all possible paths.  This has implications for how we measure the
superfluid
density $\rho_s$.  Consider the well-known relation,\cite{Pollock87}
\begin{equation}
  \label{RhoS}
  \rho_s=\frac{\bigl\langle \hat W^2\bigr\rangle}{2dt\beta}
\end{equation}
where $d$ is the dimension of the system, and $\bigl\langle\hat
W^2\bigr\rangle$ the mean square of the winding number operator which
measures the net flow of particles off the right side of the lattice
and over to the left side.  Our local moves cannot change the winding,
as is evident by comparing the configurations of Fig.~\ref{Winding}.
Hence, using Eq.~(\ref{RhoS}) our algorithm would systematically give
a zero value for the superfluid density, if we begin in a $W=0$
configuration.

However, we can still measure $\rho_s$.  The
procedure is as follows: We define the pseudo-current operator (using
the notation of Fig.~\ref{Plaquette}b for the site labels)
\begin{equation}
  \label{Pseudo-current}
  \begin{array}{l}
    \displaystyle\hat j_x(k)=\sum_{c\in(k,k+1)}\left(\hat n_{c6}+\hat
    n_{c8}-\hat n_{c2}-\hat n_{c4}\right)\\ \displaystyle\hat
    j_y(k)=\sum_{c\in(k,k+1)}\left(\hat n_{c7}+\hat n_{c8}-\hat
    n_{c3}-\hat n_{c4}\right)
  \end{array}
\end{equation}
where the sum $\sum_{c\in(k,k+1)}$ is over all cubes of
Fig.~\ref{Decomposition}b between times $k$ and $k+1$. These
quantities measure the number of particles which jump in the positive
$x$ and $y$ directions between imaginary times $k$ and $k+1$. The
winding number operators in both $x$ and $y$ directions are then given
by
\begin{equation}
  \label{WindingNumber-1}
  \begin{array}{l}
    \displaystyle\hat W_x=\frac{1}{N_x}\sum_{k=1}^M \hat j_x(k)\\
    \displaystyle\hat W_y=\frac{1}{N_y}\sum_{k=1}^M \hat j_y(k)
  \end{array}
\end{equation}
where $N_x$ and $N_y$ are respectively the number of sites in the $x$
and $y$ directions.  We measure also the auto-correlation function of
the pseudo-current:
\begin{equation}
  \label{Correlation}
  \begin{array}{l}
    \displaystyle \mathcal J_x(k)=\sum_{k^\prime}
    \left\langle\hat j_x(k+k^\prime)\hat j_x(k^\prime)\right\rangle \\
    \displaystyle \mathcal J_y(k)=\sum_{k^\prime}
    \left\langle\hat j_y(k+k^\prime)\hat j_y(k^\prime)\right\rangle
  \end{array}
\end{equation}
and compute its Fourier transform via
\begin{equation}
  \label{Fourier-1}	\widetilde{\mathcal J_{x,y}}(\omega_n)=\sum_k\mathcal J_{x,y}(k)e^{i\omega_n k}
\end{equation}
where $\omega_n=\frac{2\pi}{M}n$.\\

\begin{figure}[h!]
   \centerline{\epsfig{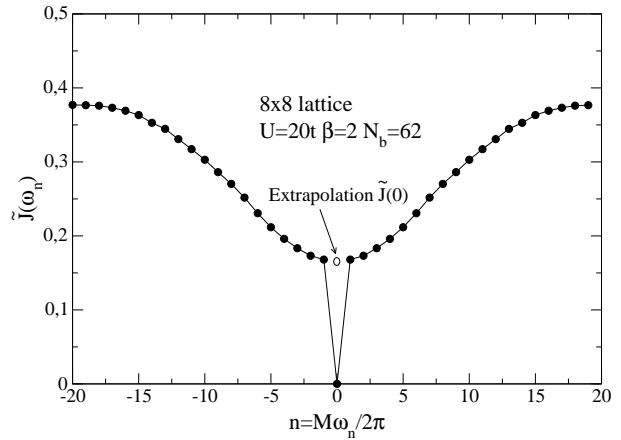}}
  \caption{\label{Extrapolation} {Restricting the winding number to a
zero value introduces a discontinuity in the Fourier transform of the
pseudo-current auto-correlation function. Performing an extrapolation
gives the correct value of the superfluid density.}}
\end{figure}

One can show that these pseudocurrent operators are related to the
winding by,
\begin{equation}
  \label{WindingNumber-2}
  \begin{array}{l}
    \displaystyle\bigl\langle\hat
    W_x^2\bigr\rangle=\frac{\widetilde{\mathcal J_x}(0)}{N_x^2}\\
    \displaystyle\bigl\langle\hat
    W_y^2\bigr\rangle=\frac{\widetilde{\mathcal J_y}(0)}{N_y^2}
  \end{array}
\end{equation}
As already stated, the measurement of the winding number, and hence
the Fourier transform at $\omega_n=0$, will give a zero value.  but we
can avoid the problem by considering the finite frequency response.
In fact, one can show\cite{Batrouni90} that the correct, nonzero,
winding number and $\rho_s$ are given by taking the zero frequency
limit of $\widetilde{\mathcal J_x}(\omega_n)$, as illustrated in
Fig.~\ref{Extrapolation}.

\section*{Results: \textit{Hard-core limit}, $V=0$}

We can work in the hard-core limit by running a soft-core code at
large $U$, or more directly by forbidding the acceptance of multiple
occupancies in our Monte Carlo moves.  This also simplifies the
computation of the matrix elements in (\ref{Partition-5}) since we can
restrict to single occupancy states (and hence diagonalize only a
$16\times16$ matrix).  Such a hard-core model is equivalent to replacing the
standard bosonic commutation rules by the hard-core rules,
\begin{equation}
  \label{Hard-core-rules}
  \begin{array}{lll}
    \left\lbrace a_i,a_i\right\rbrace=0 & \left[a_i,a_j\right]=0 &
    \forall i\neq j\\ \bigl\lbrace a_i^\dagger,a_i^\dagger
    \bigr\rbrace=0 & \bigl[a_i^\dagger,a_j^\dagger \bigr]=0 & \forall
    i\neq j\\ \bigl\lbrace a_i,a_i^\dagger \bigr\rbrace=1 &
    \bigl[a_i,a_j^\dagger \bigr]=0 & \forall i\neq j \, ,
  \end{array}
\end{equation}
where $\bigl\lbrace A,B\bigr\rbrace=AB+BA$ is an anti-commutator.  The
usual bosonic commutators between operators acting on different sites
ensure that the wave function is symmetric in the exchange of
particles, as required for bosons.

Our simulations are done mainly on $16\times16$ lattices.  We always
take $t=1$ for the hopping parameter, which thereby sets the energy
scale.  We will also set the near neighbor interaction $V=0$ until the
final results section.

In order to describe the different phases of the system, in addition to the
superfluid density $\rho_s$, we measure the Fourier transforms of the
density-density and plaquette-plaquette order correlation functions,
$S(\pi,\pi)$, $P(\pi,0)$ and $P(0,\pi)$ defined by,
\begin{eqnarray}
  \label{DensityStructureFactor}
    \!\!\!\!\!\!\!\!\!\!&& S\bigl(k_x,k_y\bigr)=\frac{1}{\bigl(N_xN_y\bigr)^2}\sum_{\vec
    r,\vec r^{\:\prime}}\bigl\langle\hat n_{\vec r} \,\, \hat n_{\vec r+\vec
    r^{\:\prime}}\bigr\rangle e^{-i\vec k\cdot\vec r}\\
  \label{PlaquetteStructureFactor}
    \!\!\!\!\!\!\!\!\!\!&& P\bigl(k_x,k_y\bigr)=\frac{1}{K^2
    \bigl(N_xN_y\bigr)^2}\sum_{p,p^\prime}\bigl\langle\hat\mathcal
    K_p\hat\mathcal K_{p^\prime}\bigr\rangle e^{-i\vec k\cdot(\vec
    r_p-\vec r_{p^\prime})}
\end{eqnarray}

\begin{figure}[h!]
  \centerline{\epsfig{figure=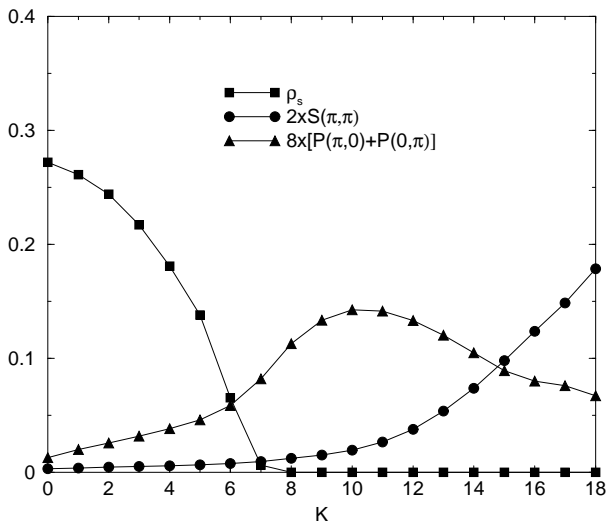,width=8cm}}
  \caption{\label{Hard-core-half-filling} {The superfluid density and
the order parameters for the hard-core case at half
filling for a $16\times16$ lattice. Increasing $K$ destroys the superfluid density. Then a solid
order appears corresponding to a VBS phase.  This phase becomes
unstable if increasing $K$ more, and yields to a CDW phase.}}
\end{figure}

Fig.~\ref{Hard-core-half-filling} shows these quantities, and the
superfluid density, as a function of $K$ at half filling. Increasing
$K$ makes the superfluid density decrease.  $\rho_s$ vanishes
completely around $K=8$. At the same time, the plaquette structure
factors $P(\pi,0)$ and $P(0,\pi)$ start to grow, indicating the
formation of stripes by the ring exchange.  This is the VBS (Valence
Bond Solid) phase, previously discussed in quantum spin systems.
Stripe formation is also observable by considering the correlations in
the hopping from plaquette to plaquette, rather than the ring
exchange.  For higher values of $K$, the VBS phase becomes unstable
and is replaced by a CDW as shown by the growth of
$S(\pi,\pi)$.\cite{Melko04}

\begin{figure}[h!]
   \centerline{\epsfig{figure=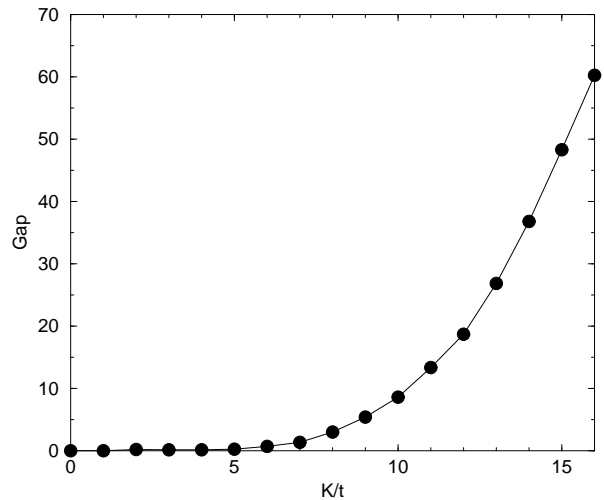,width=8cm}}
  \caption{\label{GapVsK} {The gap in the chemical potential at half
filling, as a function of $K$ for a $16\times 16$ Lattice. The gap starts to appear at $K\simeq
8$, coincident with the vanishing of $\rho_s$, and grows as $K$ is
increased, showing the presence of insulating phases. }}
\vskip 0.5cm
\end{figure}

Both the VBS and CDW phases are gapped.  That is, the energy needed to
add a particle to the lattice (the chemical potential) jumps abruptly
across $\rho=\frac12$.  We define $G=\mu_+-\mu_-$ with
$\mu_+=E\bigl(N_b = {N \over 2}+1\bigr)-E\bigl({N \over 2}\bigr)$ and
$\mu_-=E\bigl({N \over 2})-E\bigl({N \over 2}-1\bigr)$. $G=0$ in the
superfluid phase $K<8$, and is non-zero for $K>8$.  The VBS and CDW
phases are insulators.  Fig.~\ref{GapVsK} shows $G$ as one increases
$K$ out of the superfluid phase at half-filling.

\begin{figure}[h!]
   \centerline{\epsfig{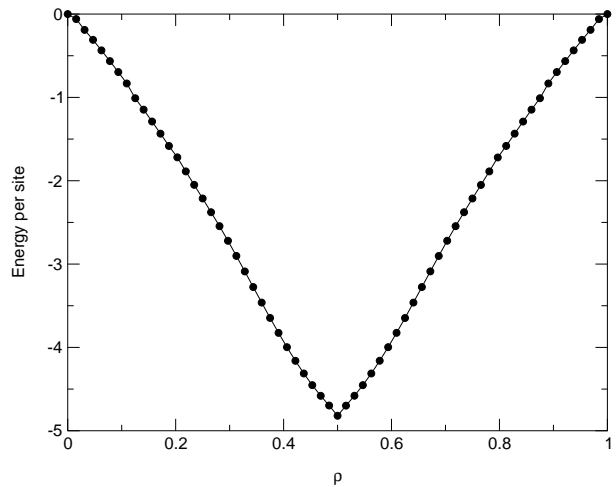}}
  \caption{\label{EnergyVersusRho} {The energy per site as a
  function of the density $\rho$, for $K=10$ on a $8\times 8$ Lattice.}}
\end{figure}

We now consider what happens when the
system is away from half-filling.  Fig.~\ref{EnergyVersusRho} shows the energy as a
function of density.  There are several interesting features.  First,
the kink in the energy right at half-filling is associated with a
nonzero gap.  Second, there is a region of densities for which the
energy is concave down.  As a consequence, the usual Maxwell
construction indicates that it is energetically more favorable to have
two separate regions of different density than a single region of
uniform density. The system phase separates.

\begin{figure}[h!]
   \centerline{\epsfig{figure=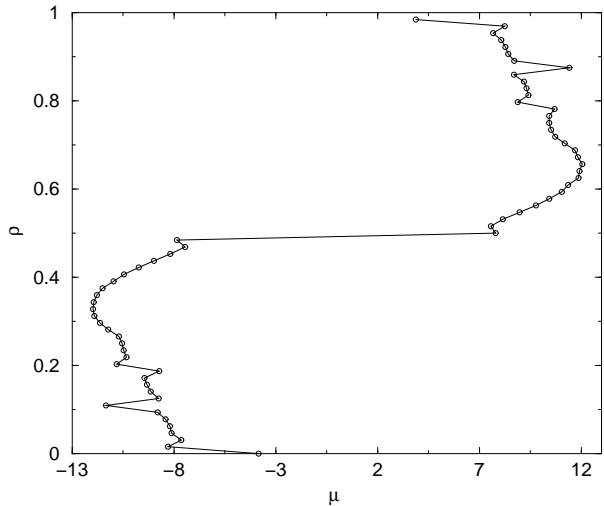,width=8cm}}
  \caption{\label{ChemicalPotential} {The density of particles as a
function of the chemical potential, for $K=10$ ($8\times 8$ Lattice).}}
\end{figure}

The energetic signature of phase separation is further illustrated in
Fig.~\ref{ChemicalPotential}, which shows the density of particles
$\rho$ as a function of the chemical potential $\mu$ over a wide range
of densities.  The jump of the chemical potential at half filling is
the gap associated with the solid VBS phase discussed above.  The
regions where the slope of the curve (proportional to the
compressibility $\kappa_T$) is negative indicate that the system is
thermodynamically unstable \cite{Batrouni00} and undergoes phase
separation.  There are two such regions.  The first is seen as soon as
we go away from half filling, ie for the two fillings $N/2 \pm 1$
immediately adjacent to half-filling, and indicates a first order
transition from the VBS phase to a superfluid.  (See also
Fig.~\ref{Hard-core-doped}.)  The same negative compressibility
near half-filling is obtained for values of $K$ higher than 15,
showing a first order transition from the CDW phase to a superfluid.
The second region of negative compressibility occurs below
$\rho_c=0.34$.

The spikes occuring at $\rho\approx 0.1$ and
$\rho\approx 0.2$ in Fig.~\ref{ChemicalPotential} are not due to statistical fluctuations. Indeed, when phase separating, the system
can form a stripe in its density profile rather than an island, due to periodic boundary
conditions. This produces a shift in the energy,
and hence in the chemical potential. In our simulations,
the system forms a stripe for $\rho\in[0.1,0.2]$,
and an island when crossing the limits, generating the
spikes.

\begin{figure}[h!]
   \centerline{\epsfig{figure=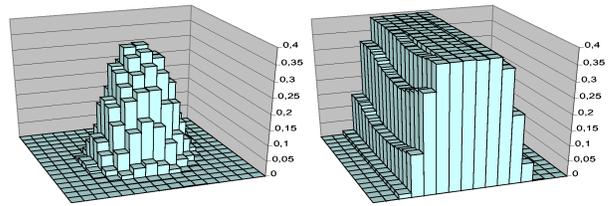,width=8cm}}
  \caption{\label{TypicalDensityProfiles} {Typical QMC results for the
average density distribution when the clustering occurs. For a density
$\rho=12/256$ the system adopts a shape of an island, while for
$\rho=48/256$ the system prefers to form a stripe, which, in the
presence of periodic boudary conditions, is the geometry which
maximizes the number of occupied four site plaquettes for which the
ring exchange energy is low.}}
\end{figure}

\begin{figure}[h!]
  \centerline{\epsfig{figure=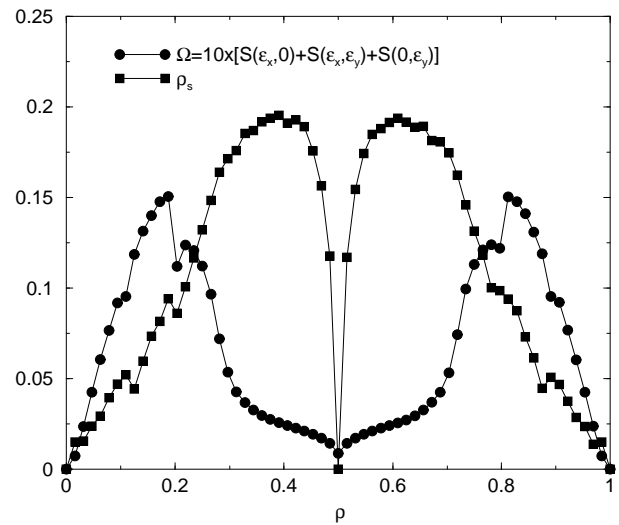,width=8cm}}
  \caption{\label{Hard-core-doped} {Superfluid density and $\Omega$ as
functions of the density of particles, for $K=10$ on an $8\times 8$ Lattice.}}
\end{figure}

While analysis of the energy and chemical potential yield quantitative
information about the regions of stability, a more straightforward,
qualitative, indication of clustering comes from simple real space
images of the boson density during the course of a simulation.
Whether clustering occurs or not, if the density is averaged over the
whole simulation, we should observe a uniform density distribution,
since the probability of a state is independant of any translation in
the lattice. But if the density is averaged over only a few
iterations,
we observe a density profile reflecting the clustering
(Fig.~\ref{TypicalDensityProfiles}).

To corroborate further that this
negative compressibility is associated with phase
separation, we define the observable $\Omega$ to be the average of the
density structure factor over the smallest values of the wave vector:
\begin{equation}
  \label{OmegaDetector}
  \Omega=\frac{S(\epsilon_x,0)+S(\epsilon_x,\epsilon_y)+S(0,\epsilon_y)}3
  \hspace{0.5cm} \epsilon_{x,y}=\frac{2\pi}{N_{x,y}}
\end{equation}

$\Omega$ measures modulations of the density profile with wave lengths
on the order of the size of the lattice. Fig.~\ref{Hard-core-doped}
shows the superfluid density and $\Omega$ as functions of the density
of particles $\rho$, for a fixed value $K=10$. The symmetry of the
curves reflects the particle-hole symmetry of the hard-core
Hamiltonian. Starting from $\rho=\frac 12$ where $\Omega$ and $\rho_s$
have a zero value, and which corresponds to the VBS phase, we see that
doping the system increases the superfluid density rapidly.  $\rho_s$
reaches its maximum for $\rho\simeq 0.4$ (or $\rho\simeq 0.6$ by
symmetry). Then $\rho_s$ falls at the same time that $\Omega$ grows
and reaches its maximum for $\rho\simeq 0.2$ (or $\rho\simeq 0.8$).
Fig.~\ref{Hard-core-doped} suggests that as soon as we go away from
half-filling, there is a transition from the VBS phase first to a
stable superfluid, but that subsequently phase separation sets in.
The density $\rho \simeq 0.3$ where $\rho_s$ falls and $\Omega$
increases rapidly matches quite well with the density at which
$\kappa$ changes sign in Fig.~\ref{ChemicalPotential}.

The occurrence of phase separation indicated in the
compressibility, $\Omega$, and in real space images, may be understood
with a simple physical picture. When $K$ is sufficiently large, the
system increases the ring exchange processes in order to decrease its
energy.  But ring exchange is possible only if there are second
neighbor particles.  Thus in a dilute collection of particles, the
ring exchange term has a tendency to act as an attractive potential.
This is not true of the usual kinetic energy, since there a single
boson can hop by itself without needing a partner.

\begin{figure}[h!]
   \centerline{\epsfig{figure=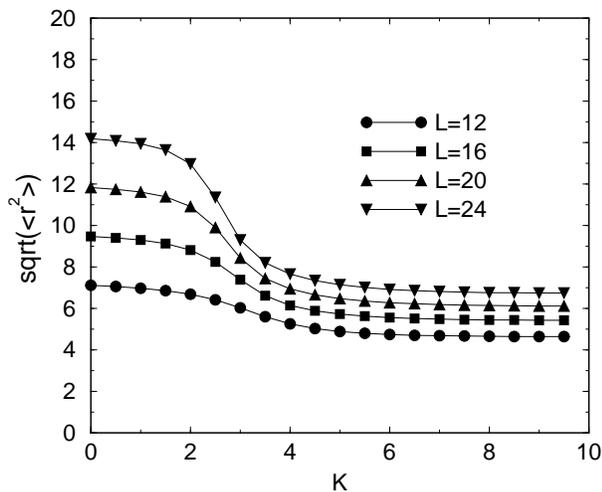,width=8cm}}
  \caption{\label{TwoBosons} {The mean square distance
$\sqrt{\bigl\langle\hat r^2\bigr\rangle}$ between two particles as a
function of $K/t$, in the soft-core case with $U=12t$ at $T=0$.}}
\end{figure}

One can analyze this binding quantitatively for the case of two bosons
by exact diagonalization (Lanczos).  Fig.~\ref{TwoBosons} shows the
square root of the mean quadratic distance $\bigl\langle\hat
r^2\bigr\rangle$ between two particles for different $L\times L$
lattice sizes in the ground state.  The soft-core case is considered
with an on-site repulsion $U=12t$.  For low values of $K$ the distance
between the bosons grows linearly with $L$, as one would expect if the
particles were distributed independently across the lattice.  On the
other hand, for high values of $K$, the distance between the particles
is relatively independent of lattice size, suggesting the particles
stay together, and demonstrating the existence of a bound state.
While the data shown are for $U=12t$, the phenomenon is
independent of $U$, and, in particular, is seen also in the hard-core
case.
The reason the on-site repulsion $U$ does not successfully compete
with the binding is that the bound wavefunction favors the bosons on
second-near-neighbor sites where $U$ plays no role.  This is reflected
in the data of Fig.~\ref{TwoBosons} where we see that the distance
between the paired bosons is still several lattice sites.

\begin{figure}[h!]
   \centerline{\epsfig{figure=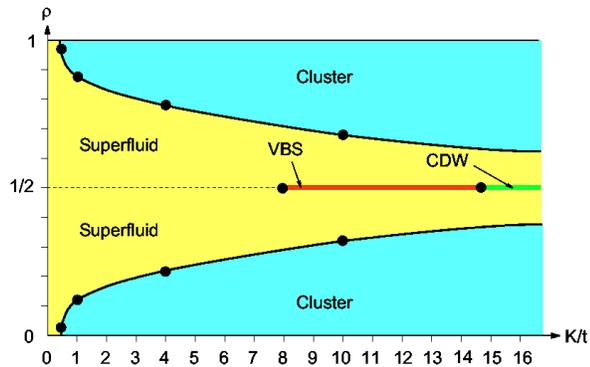,width=8cm}}
  \caption{\label{Hard-core-phase-diagram} {The phase diagram of
hard-core bosons ($16\times 16$ Lattice). The solid phases (VBS and CDW) exist only at
half-filling.  When doping the system, the phase first becomes
superfluid, then the clustering occurs closer to $\rho=\frac12$ as $K$
is increased.}}
\end{figure}

Figs.~\ref{ChemicalPotential} and \ref{Hard-core-doped} showed our
results for hard-core bosons at fixed $K=10$ as the system is doped.
By performing further simulations for different values of $K$, we are
able to generate the complete phase diagram of hard-core bosons
(Fig.~\ref{Hard-core-phase-diagram}) in the $K-\rho$ plane.  Gapped
VBS and CDW phases exist only at half-filling and relatively large
$K$.  For weaker $K$ and half-filling the system is superfluid.  When
the system is doped a small amount away from half-filling, only a
superfluid phase is present.  Upon further doping, the clustering
region is reached.  As $K$ becomes large, clustering occurs closer and
closer to half-filling.

Melko \textit{et al.} \cite{Melko04} have studied the spin-1/2 Hamiltonian,
\begin{equation}
  \label{Melko-1}	\hat\mathcal H=-J\sum_{\bigl\langle
    ij\bigr\rangle}\hat\mathcal B_{ij}-K\sum_{\bigl\langle
    ijkl\bigr\rangle}\hat\mathcal P_{ijkl}-h\sum_i\hat\mathcal S_i^z
\end{equation}
where
\begin{equation}
  \label{Melko-2}
\begin{array}{ll}
\hat\mathcal B_{ij}=2\bigl(S_i^xS_j^x+S_i^yS_j^y\bigr)\\ \hat\mathcal
P_{ijkl}=S_i^+S_j^-S_k^+S_l^-+S_i^-S_j^+S_k^-S_l^+
\end{array}
\end{equation}
and $S^x$, $S^y$, $S^z$, $S^+$, $S^-$ are the usual spin
operators. This is the Heisenberg $XY$ model with a four spin exchange
term and in a magnetic field. This model is exactly equivalent to our
bosonic model in the hard-core limit. The magnetic field plays the
role of the chemical potential in the grand canonical ensemble. In the
phase diagram obtained by Melko {\it et al.}, the superfluid, VBS and
CDW phases are present and their boundaries match with our results. In
addition the nature of the transitions also agree. Our simulations in
the canonical ensemble exhibit clear phase separation indicating first
order transitions from CDW insulator (N\'eel) to SF and also from SF
to Mott (fully polarized, in the spin language). To see this, consider
the phase diagram, Fig.~1 in Ref.~[7]. As discussed in that
paper, the cuts along lines $A$ and $B$ show clear hysteresis
indicating first order transitions. Our simulations in
Fig.~\ref{ChemicalPotential} are done at constant $K$ and would
correspond most closely to a vertical cut, like $A$, in
Ref.~[7] but going from N\'eel to SF to fully
polarized. This way two first order phase transitions would be crossed,
as shown in Fig.~\ref{ChemicalPotential} here by the negative
compressibility regions. Note also that the jump in the density when
going from N\'eel to SF (e.g. cut $B$ in~[7]) is very small
as shown in the magnetization histogram, Fig.~3
of~[7]. Recalling that $\rho = m +0.5$, we see that the jump
in $\rho$ in the histogram is about $0.015$, of the same order as we
see in our Fig.~\ref{ChemicalPotential}.

Thus, the hard-core bosonic model with ring exchange does not exhibit
any compressible but non-superfluid, Bose-liquid, phase.  In the next
section we examine the possibility\cite{Paramekanti02} that the
relaxation of the hard-core constraint could make the solid phases
(VBS or CDW) evolve to a Bose liquid.

\section*{Results: \textit{Soft-core case}, $V=0$}

\begin{figure}[h!]
   \centerline{\epsfig{figure=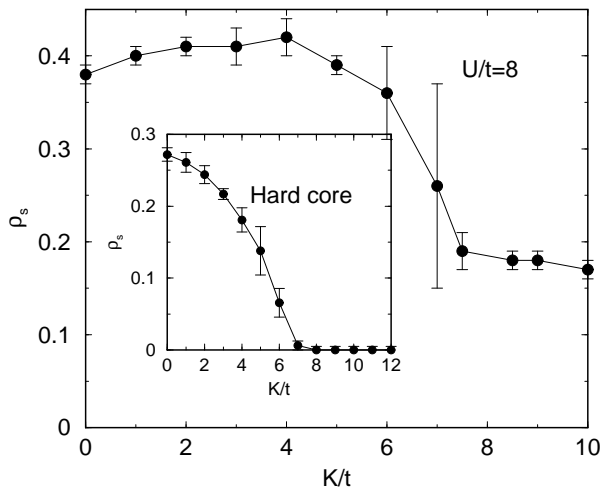,width=8cm}}
  \caption{\label{RhosVsK-Soft-core} {The superfluid density as a
function of $K$ for the soft-core case where $U=8t$, at half-filling ($16\times 16$ Lattice).
$\rho_s$ does not vanish when $K$ is large, unlike the hard-core
case.}}
\end{figure}

We now turn to the soft-core case, but still choose the intersite
repulsion $V=0$ and half-filling.  The superfluid density $\rho_s$ is
shown as a function of $K$ for $U=8t$ in
Fig.~\ref{RhosVsK-Soft-core}.  $\rho_s$ first grows slightly when $K$
increases.  When $K\simeq 4t$, $\rho_s$ starts to decrease.  This
decrease becomes rather rapid until, at $K\simeq 8t$, $\rho_s$ levels
off at about half its $K=0$ value.  This behavior is quite different
from the hard-core case (inset) where $\rho_s$ decreases as soon as
the ring exchange interaction is turned on, and {\it vanishes} at
$K\simeq 8t$.

\begin{figure}[h!]
   \centerline{\epsfig{figure=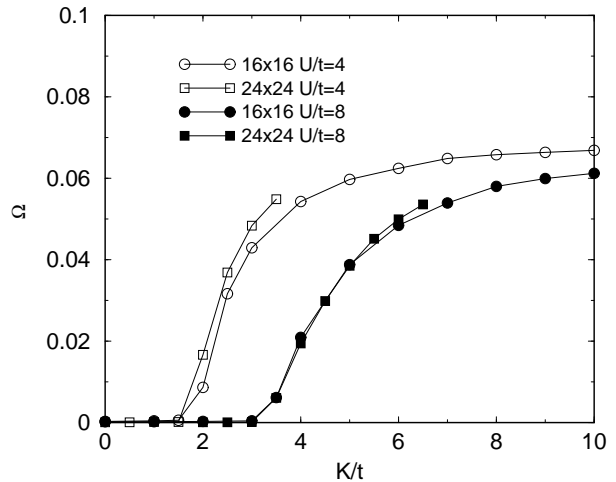,width=8cm}}
  \caption{\label{OmegaVsK-Soft-core} {$\Omega$ versus $K$ for $U=4t$
and $U=8t$ and two different lattice sizes. The lattice is
half-filled.  We can see that $\Omega$ increases sharply at $K\simeq
U/2$.}}
\end{figure}

In order to understand the difference in the behavior of $\rho_s$
between the soft and hard core cases, we begin by examining $\Omega$,
the small momentum density-density structure factor.
Fig.~\ref{OmegaVsK-Soft-core} shows $\Omega$ as a function of $K$ for
$U=4t$ and $U=8t$.  In both cases, when $K$ reaches a value of the
order $U/2$, $\Omega$ grows sharply, showing, as in the hard-core case,
the presence of a clustering tendency.  Results for $16\times16$
and $24\times24$
lattices are almost identical, indicating that the phase separation is
not a finite size effect.

The behavior of the superfluid density can then be explained as
follows: When $K$ is weak, the ring exchange term supports the
motion of the particles, and helps the kinetic term in delocalizing
them. A flow can thus occur more easily and $\rho_s$ increases.  When
$K$ is large, clustering occurs and long range flow across the lattice
is somewhat inhibited.  If, for example, the system forms a stripe,
one might argue that roughly speaking the flow of particles is
possible only in one direction, and $\rho_s$ should decrease by a
factor of two, in agreement with Fig.~\ref{RhosVsK-Soft-core}.

\begin{figure}[h!]
\centerline{\epsfig{figure=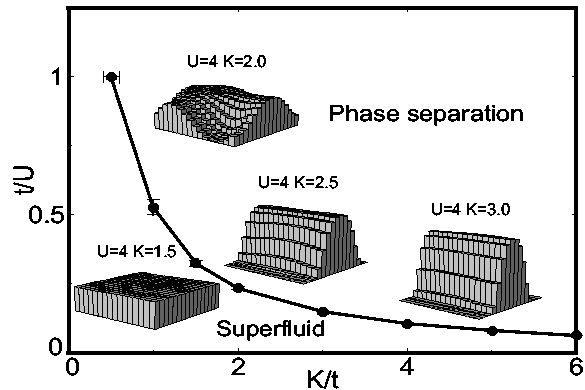,width=8cm}}
  \caption{\label{Soft-core-phase-diagram} {The phase diagram of
  soft-core bosons ($16\times 16$ Lattice). The phase is superfluid under the curve, while the
  system forms a cluster in the upper part. The curve is approximately
  hyperbolic.}}
\end{figure}

As we have seen, at half-filling, in the hard-core case, superflow
does not occur when $K$ is large because of the formation of gapped
VBS and checkerboard phases.  For the soft-core case, no solid
structure is established and the particles can circulate. Thus the
soft-core model is always superfluid at half filling and undergoes a
clustering when $K$ reaches a value of the order $U/2$.
Fig.~\ref{Soft-core-phase-diagram} shows the phase diagram of
soft-core bosons at half filling.

\begin{figure}[h!]
\centerline{\epsfig{figure=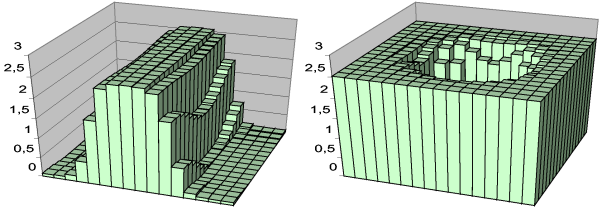,width=8cm}}
  \caption{\label{Soft-core-Density-Profiles} {Density profiles for the
  soft-core case with $U=4$ and $K=8$ for commensurate fillings, $\rho=1$
  (left) and $\rho=2$ (right).}}
\end{figure}

Our phase diagram Fig.~\ref{Soft-core-phase-diagram} is for
half-filling, but the main feature, a tendency to phase separation, is
the same at other densities.  As an extreme example, simulations done
for a commensurate filling of one particle per site show that, like
the superfluid phase, the Mott phase also collapses when the ring
exchange processes are too strong, and is replaced by clustering
(Fig.~\ref{Soft-core-Density-Profiles}).

\section*{Results: \textit{Soft-core case}, $V \neq 0$}

\begin{figure}[h!]
   \centerline{\epsfig{figure=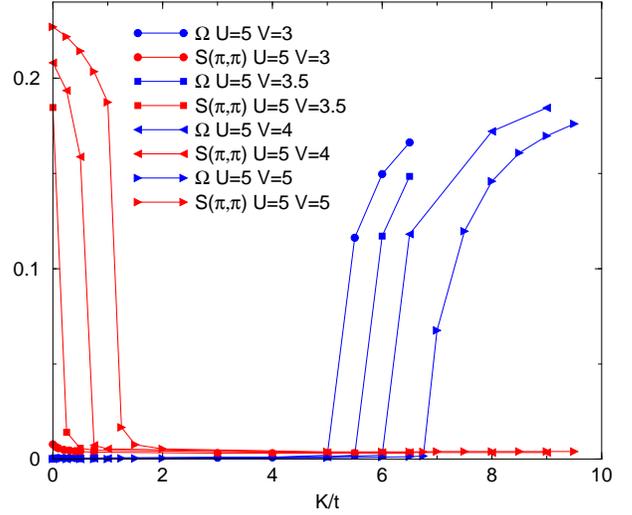,width=8cm}}
  \caption{\label{OmegaAndSpipiVsK} {The CDW structure factor
$S(\pi,\pi)$ and $\Omega$ for $U=5t$ and different values of the
potential $V$ between first nearest neighbors, at half filling ($16\times 16$ Lattice).  $V$
is seen to compete with phase separation, pushing the onset of the
rise in $\Omega$ out to larger and larger $K$.}}
\end{figure}

\begin{figure}[h!]
   \centerline{\epsfig{figure=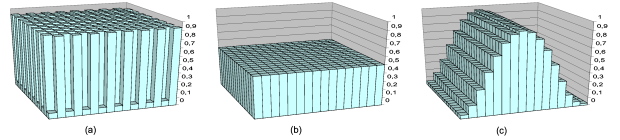,width=8cm}}
  \caption{\label{First-nearest-neighbors} {Density profiles at half
filling for $U=5t$, $V=5t$, and different values of $K$: (a) $K=t$,
CDW phase; (b) $K=5t$, uniform density; and (c) $K=7t$, phase
separation.}}
\end{figure}

In the preceding two sections we have seen that a Bose-liquid does not
occur in either the hard- or soft-core Bose Hubbard models when ring
exchange is included, in contrast to recent
suggestions.\cite{Paramekanti02} It is natural to consider whether
longer range repulsion might prevent the system from collapsing.  As a
first step, we now include a repulsion between first nearest
neighbors.  Fig.~\ref{OmegaAndSpipiVsK} shows the structure factor
$S(\pi,\pi)$ and $\Omega$ for $U=5t$ and different values of $V$, as a
function of $K$ at half-filling. Three different behaviors of the
density correlations are observable: a CDW (density order at momentum
$(\pi,\pi)$) when $K$ is weak (well known in the model without ring
exchange processes),\cite{Niyaz93} at intermediate $K$ a uniform phase
where the density structure factor is small at all wavelengths,
and a regime of phase separation ($\Omega$ large) when $K$ is big.  As
expected, $V$ does suppress the tendency for clustering.  $K$ must be
made larger for phase separation to occur when $V$ is increased.  The
phases of Fig.~\ref{OmegaAndSpipiVsK} are also directly visible in
real space snapshots of the densities in the course of the simulations
(Fig.~\ref{First-nearest-neighbors}).

\begin{figure}[h!]
   \centerline{\epsfig{figure=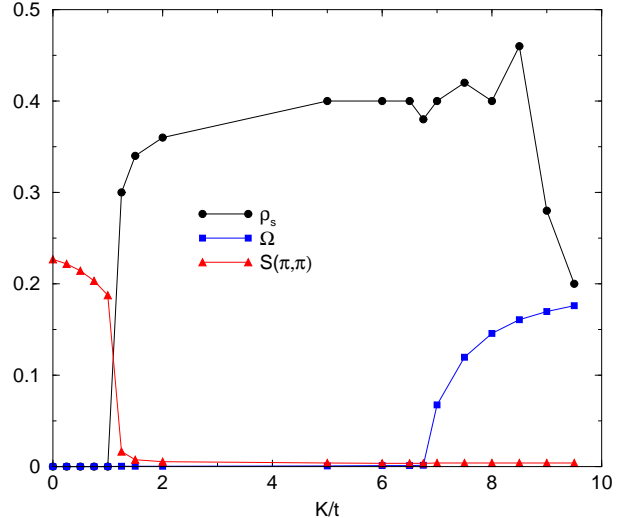,width=8cm}}
  \caption{\label{RhosVsK-First-nearest}        {The superfluid density
$\rho_s$, the structure factor
  $S(\pi,\pi)$, and $\Omega$ as functions of $K$ for $U=5t$ and $V=5t$ ($16\times 16$ Lattice).
The superfluid density vanishes only when the system is in the CDW phase.}}
\end{figure}

Despite the absence of density order, the uniform phase
(Fig.~\ref{First-nearest-neighbors}b) is not a Bose liquid.
Fig.~\ref{RhosVsK-First-nearest} shows that the superfluid density is
zero only for the CDW phase. As soon as the staggered density order is
destroyed with increasing $K$, the superfluid density becomes nonzero.
$\rho_s$ climbs rapidly and is roughly constant through the region
where both $S(\pi,\pi)$ and $\Omega$ are small.  It remains non-zero
in the cluster phase, though it does drop by a factor of two when
clumping begins.  There is no Bose liquid phase in the model.
(Of course, if the density is sufficiently small, then an island phase in
which the boson cloud does not span the system will have vanishing
$\rho_s$.  This situation also occurs for bosons in a confining potential,
where the phase is nevertheless commonly referred to as being superfluid.

\section*{Conclusions}

In this paper we have studied the effect of a ring exchange term in
the hard- and soft-core Bose Hubbard model, also including a
near-neighbor repulsion, using Quantum Monte Carlo simulations in the
canonical ensemble.  It had been suggested that this term might lead
to a normal Bose liquid phase, that is, one which is compressible but
non superfluid.  For the Hard-core case, we reproduced results
obtained by Melko \textit{et al.} \cite{Melko04} in the grand
canonical ensemble.  However, working in the canonical ensemble
enables us to capture and characterize a hitherto unsuspected cluster
phase of the model.  No Bose liquid phase was observed.  Thus the
speculation of Paramekanti \textit{et al} \cite{Paramekanti02}, that
the relaxation of the hard-core constraint might give rise to a Bose
liquid does not appear to be borne out.  The tendency to clustering
can be understood by analyzing the effect of $K$ on a two boson
system.  We have shown numerically that $K$ acts much like an
attractive potential, and that when it is sufficiently large, the two
boson form a bound state.  This attraction leads to clustering in
systems with larger boson numbers.

We saw that, while a near neighbor repulsion competes with this
clustering, it could not create a normal bose liquid.  At weak $K$ it
promotes an alternate charge density wave insulator.  At strong $K$,
while $V$ does help prevent clustering, $\rho_s$ remains nonzero.  It
might be interesting to study models with longer range interactions
which will act together to prevent phase separation, but compete with
each other by promoting phases at different ordering momenta.  Such
models might, finally, realize the Bose liquid phase.

 To conclude, let us comment on the implications of our work for the
 phase diagram of the spin-1/2 quantum Heisenberg model with a ring
 exchange term.  The kinetic energy term in the hard-core boson
 Hubbard model maps onto $ J \sum (S_{{\bf i}}^{x} S_{{\bf j}}^{x} +
 S_{{\bf i}}^{y} S_{{\bf j}}^{y})$ with exchange constant $J=2t$.  At
 the value $U=4t$ in our soft core model, double occupancy is already
 very rare at half-filling, and hence we are almost in the hard-core
 limit.  The value of the ring exchange energy scale required to drive
 phase separation for this $U$ is $K \approx 2t$, or in other words,
 $K \approx J$.  To reproduce the near-neighbor coupling of the $z$
 components of spin in the Heisenberg model we must include a
 near-neighbor repulsion in the bose-Hubbard model, a term which
 clearly would suppress phase separation.  Thus we expect ring
 exchange to have the potential to drive phase separation in the
 Heisenberg model only for $K$ considerably greater than $J$.

We acknowledge support from the National Science Foundation under
awards NSF DMR 0312261 and NSF INT 0124863, and useful input
from R. Stones.


\begin{thebibliography}{10}
\bibitem{twositedecoupling}
See for example, reference \cite{Batrouni90}
where only two site decouplings were used.
\bibitem{highocc}If one were interested in cases where the density
were larger, and the on-site $U$ were small, this might no longer be a
good approximation.
\end{thebibliography}
\end{document}